\documentclass[traditabstract]{aa}

\usepackage{natbib}
\bibpunct{(}{)}{;}{a}{}{,} % to follow the A&A style
\usepackage{amssymb}
\usepackage{url}
\usepackage{graphicx}
\usepackage{longtable}
\usepackage[usenames,dvipsnames]{color}

\newcommand{\submm}{submillimetre}
\defcitealias{michalowski10smg}{M10} 	
\defcitealias{hainline11}{H11} 	
\defcitealias{cirasuolo10}{C10} 
\newcommand{\mich}{\citetalias{michalowski10smg}}
\newcommand{\hain}{\citetalias{hainline11}}
\newcommand{\ciras}{\citetalias{cirasuolo10}}

\begin{document}

 \title{The stellar masses and specific star-formation rates of submillimetre galaxies}
 
\titlerunning{The stellar masses and specific star-formation rates of submillimetre galaxies}
\authorrunning{Micha{\l}owski et al.}

\author{Micha{\l}~J.~Micha{\l}owski\inst{1}
\and
James~S.~Dunlop\inst{1}
\and
Michele Cirasuolo\inst{1}
\and 
Jens~Hjorth\inst{2}
\and
Christopher~C.~Hayward\inst{3}
\and
Darach~Watson\inst{2}
	}

\institute{
SUPA\thanks{Scottish Universities Physics Alliance}, Institute for Astronomy, University of Edinburgh, Royal Observatory, Edinburgh, EH9 3HJ, UK, {\tt mm@roe.ac.uk}
\and
Dark Cosmology Centre, Niels Bohr Institute, University of Copenhagen, Juliane Maries Vej 30, 2100 Copenhagen \O, Denmark
\and
Harvard-Smithsonian Center for Astrophysics, 60 Garden St., Cambridge, MA 02138, USA
}

%----------------------------------ABSTRACT----------------------------------
\abstract{Establishing the stellar masses, and hence specific star-formation rates 
of {\submm} galaxies is crucial for determining the role of such objects 
in the cosmic history of galaxy/star formation. However, there is as yet 
no consensus over the typical stellar masses of {\submm} galaxies, as 
illustrated by the widely differing results reported from recent optical-infrared 
studies of {\submm} galaxies with spectroscopic redshifts $z \simeq 2$--$3$.
Specifically, even for the same set of {\submm} galaxies, 
the reported average stellar masses have ranged over an order of magnitude, 
from $\simeq 5 \times 10^{10} {\rm \,M_{\odot}}$ to $\simeq 5 \times 10^{11} {\rm \,M_{\odot}}$. 
Here we study how different methods of analysis 
can lead to such widely varying results. We find that, 
contrary to recent claims in the literature, potential contamination of  
IRAC $3$--$8$ $\mu$m photometry from hot dust associated with an active nucleus 
is not the origin of the published discrepancies in derived stellar masses.
Instead, we expose in detail how inferred stellar mass depends 
on assumptions made in the photometric fitting, 
and quantify the individual and 
cumulative effects of different choices of initial mass function, 
different ``brands'' of evolutionary synthesis models, and different forms 
of assumed star-formation history.
We review current observational evidence for and against 
these alternatives as well as clues from the hydrodynamical simulations, 
and conclude that, for the most justifiable choices of these model
inputs, the average stellar mass of luminous ($S_{850} \gtrsim 5$\,mJy)
{\submm} galaxies is $\simeq 2 \times 10^{11} {\rm \,M_{\odot}}$ to within a factor $\simeq 2$. 
We also check and confirm that this number is perfectly reasonable in the light of the latest measurements
of the dynamical masses of these objects ($\simeq 2-6 \times 10^{11} {\rm \,M_{\odot}}$ from CO (1-0) observations), 
and the evolving stellar mass function of the overall galaxy population. Galaxy stellar masses of 
this order imply that the average specific star-formation rate of {\submm} 
galaxies is comparable to that of other star-forming galaxies at $z > 2$, at
$2$--$3$\,Gyr$^{-1}$.
This supports the view that, while rare outliers may be found at any stellar mass, most {\submm} galaxies 
simply form the top end of the `main-sequence' of star-forming galaxies
at these redshifts. Conversely, this argues strongly against the viewpoint, frequently 
simply asserted in the literature, that {\submm} galaxies are extreme 
pathological objects, of little relevance in the cosmic history of star-formation.  
}


\maketitle

%----------------------------------INTRODUCTION----------------------------------
\section{Introduction}
\label{sec:intro}

The properties of galaxies selected at {\submm} wavelengths (SMGs)
have yet to be fully understood. In particular, there is still no real consensus over their stellar masses. 
A robust measurement of the stellar masses of SMGs is key to addressing a number of important issues, including  
the evolution of the relationship between galaxy and black-hole masses \citep[][\citetalias{hainline11} thereafter]{borys05,hainline11}, 
and the contribution of SMGs to the overall history of stellar mass assembly \citep[][\citetalias{michalowski10smg} hereafter]{michalowski10smg}. 
Perhaps most importantly, a proper understanding of the stellar masses of SMGs is required to distinguish whether their massive dust-enshrouded  
star-formation rates are fuelled primarily by mergers, or by cold gas infall \citep{swinbank08,dave10,gonzalez11,narayanan10,ricciardelli10,hayward11b,hayward11,hayward12}. Unless
the uncertainty in stellar masses can be reduced, it is in essence impossible to tell whether SMGs have specific star-formation rates ($\mbox{sSFR}=\mbox{SFR}/M_*$)
comparable to those of other star-forming galaxy populations at $z > 2$, or whether they are extreme outliers from the `main-sequence' of 
star-forming galaxies \citep{daddi07,noeske07,gonzalez10}.

Most previous studies of the stellar masses of SMGs have concluded in favour of values
in the range $\simeq10^{11}$--$10^{12}\,{\rm M_\odot}$ \citep{borys05,dye08}. 
However, two recent studies of $\simeq70$ SMGs with spectroscopic redshifts $\simeq2$--$3$ from \citet[][]{chapman05} have reached apparently very 
different conclusions, highlighting the impact of alternative methods of analysis. 
Specifically, \citet{michalowski10smg, michalowski10smg4} found stellar masses consistent with previous studies, with a median 
value $M_*=3.5\times10^{11}\,{\rm M_\odot}$ \citep[see also][]{hatsukade10,ikarashi11,santini10,tamura10}, whereas \citet{hainline11} have reported 
values, based on {\it the same} SMGs, which are systematically lower by a factor of $\simeq6$
\citep[median $M_*=5$--$7\times10^{10}\,{\rm M_\odot}$; see also][]{wardlow11}.

The primary objective of this paper is to explore the origins of this discrepancy, and to attempt to resolve it. More 
generally, however, we have taken this opportunity to expose in detail how the derived stellar masses (and hence sSFRs) 
of SMGs depend on different assumed forms of star-formation history, and different ``brands'' of evolutionary synthesis models.
The use of a sample of SMGs with spectroscopic redshifts is especially useful for this purpose, since it is well known
that, if only photometric redshifts are available, the errors in derived stellar masses can be dominated by the 
effect of uncertainties in $z_{phot}$.

The structure of this paper is as follows.
In Section~\ref{sec:mstar} we delineate the various factors and choices of analysis methods which influence
the stellar mass determinations and quantify their impact on the final derived values. 
We also review recent evidence that informs the preferred choice of, for example, initial mass function (IMF) and 
evolutionary synthesis model, and conclude that for most reasonable choices of such model inputs the 
derived stellar masses in fact only differ by a factor $\simeq 2$.
In Section~\ref{sec:check} we investigate the consistency of our re-derived stellar masses with 
the latest measurements
of the dynamical masses of these objects, the evolving stellar
mass function of the overall galaxy population, and the predictions of theoretical simulations.
We discuss the implications of our results on the sSFR of SMGs in 
Section~\ref{sec:SSFR}, and close with a summary of our conclusions in Section~\ref{sec:conclusion}.
We use a cosmological model with $H_0=70$\,km\,s$^{-1}$\,Mpc$^{-1}$,  $\Omega_\Lambda=0.7$ and $\Omega_m=0.3$.

%-------------------------------------------------------------------------------------------
%----------------------------------DERIVE MSTAR----------------------------------
%-------------------------------------------------------------------------------------------
\section{Determining the stellar masses of SMGs}
\label{sec:mstar}

\begin{table*}
\caption{Comparison of stellar masses and specific star formation rates of SMGs calculated using various methods, stellar population models and star formation histories. \label{tab:mcomp}   }
\centering
\begin{tabular}{lllccccccc}
\hline\hline
Fit\tablefootmark{a} & SSP\tablefootmark{b} & SFH\tablefootmark{c} & Mean $M_*$ & Median $M_{*}$ & Mean SSFR & Median SSFR & Age$_{\rm Y}$\tablefootmark{d} & Age$_{\rm O}$\tablefootmark{e} & Frac$_{\rm O}$\tablefootmark{f} \\
                     &                      &                      & ($\log M_\odot$) & ($\log M_\odot$) & (Gyr$^{-1}$) & (Gyr$^{-1}$) & (Gyr) & (Gyr) & \\
\hline
\citetalias{michalowski10smg} & Padova & cont+burst & $11.32 \pm 0.05$ & $11.31
^{+0.07}_{-0.06}$ & $2.86\pm0.43$ & $1.74^{+0.17}_{-0.46}$ & 0.050 & 1.856 & 
0.84\\
\citetalias{hainline11} & BC03 & 1burst/const & $10.87 \pm 0.06$ & $10.88^{+0.11
}_{-0.07}$ & $9.71\pm1.99$ & $4.60^{+1.22}_{-0.92}$
& 0.820 & $\cdots$ & $\cdots$ \\
\citetalias{hainline11} & M05 & 1burst/const & $10.71 \pm 0.06$ & $10.74^{+0.08
}_{-0.05}$ & $12.96\pm2.17$ & $6.85^{+1.56}_{-2.09}$
& 0.860 & $\cdots$ & $\cdots$ \\
\citetalias{cirasuolo10} & BC03 & double & $11.44 \pm 0.08$ & $11.47^{+0.11}_{-
0.10}$ & $10.87\pm8.68$ & $1.35^{+0.21}_{-0.31}$ & 0.191 & 1.332 & 0.87\\
\citetalias{cirasuolo10} & BC03 & tau & $11.16 \pm 0.10$ & $11.18^{+0.11}_{-0.05
}$ & $17.36\pm11.98$ & $2.56^{+1.00}_{-0.71}$ & 1.178 & $\cdots$ & $\cdots$ \\
\citetalias{cirasuolo10} & BC03 & 1burst & $11.06 \pm 0.10$ & $11.02^{+0.06}_{-
0.06}$ & $14.88\pm8.69$ & $3.07^{+1.16}_{-0.47}$ & 0.246 & $\cdots$ & $\cdots$ 
\\
\citetalias{cirasuolo10} & M05 & double & $11.73 \pm 0.06$ & $11.72^{+0.03}_{-
0.12}$ & $1.25\pm0.19$ & $0.67^{+0.25}_{-0.11}$ & 0.075 & 1.777 & 0.66\\
\citetalias{cirasuolo10} & M05 & tau & $10.46 \pm 0.13$ & $10.78^{+0.07}_{-0.17
}$ & $167.34\pm47.42$ & $5.16^{+6.38}_{-0.78}$ & 0.767 & $\cdots$ & $\cdots$ \\
\citetalias{cirasuolo10} & M05 & 1burst & $11.24 \pm 0.07$ & $11.22^{+0.12}_{-
0.04}$ & $8.43\pm4.57$ & $1.90^{+0.49}_{-0.43}$ & 0.198 & $\cdots$ & $\cdots$ \\
\hline
\end{tabular}
\tablefoot{ 
The \citet{chabrier03} IMF is assumed. Errors are standard deviations for the means and bootstrap $68$\% ranges for medians.
\tablefoottext{a}{SED fitting method presented in \citetalias{michalowski10smg} \citep{michalowski10smg}, \citetalias{hainline11} \citep{hainline11}, or \citetalias{cirasuolo10} \citep{cirasuolo10}.}
\tablefoottext{b}{Single stellar population models being either Padova tracks, BC03 \citep{bruzualcharlot03} or M05 \citep{maraston05}.}
\tablefoottext{c}{Assumed star formation history being either a continous SFH with a burst (cont+burst), an average of the single burst and constant (1burst/const), double burst (double), single burst (1burst), or exponentially declining (tau).}
\tablefoottext{d}{Mean age of the younger stellar component.}
\tablefoottext{e}{Mean age of the older stellar component.}
\tablefoottext{f}{Mean fractional contribution to the total stellar mass of the older component.}
}
\end{table*}

%-----------------IMF-----------
\subsection{Initial mass function}
\label{sec:imf}

The choice of a particular IMF introduces a systematic uncertainty of a factor of $\simeq 2$ in the determination of the stellar masses ($M_*$) 
and star-formation rates  \citep[SFRs;][]{erb06}. Specifically, the differences between the IMFs assumed by {\mich} \citep{salpeter} and {\hain} \citep{chabrier03} accounts for a 
factor of $1.8$ in the difference between the quoted stellar mass estimates 

Both top-heavy and standard  IMFs have been claimed to reproduce the number counts of SMGs \citep{baugh05,fontanot07,hayward11}. 
{ This confusing situation is further exacerbated by the fact that different studies of local ellipticals (arguably the likely present-day 
descendants of SMGs) have yielded a wide range of preferred IMFs including bottom-light \citep{vandokkum08}, \citet{kroupa01} \citep{gerhard01,cappellari06},    
Salpeter \citep{grillo09}, and even bottom-heavy (\citealt{vandokkum10, vandokkum11}; see also \citealt{thomson11}).}

Hence, since there is as yet no clear evidence that the IMF of SMGs is 
systematically different from that of other galaxies, in this paper we assume the \citet{chabrier03} IMF (unless stated otherwise) to which all 
appropriate derived quantities are converted if necessary { (values derived using the \citealt{salpeter} IMF are divided by $1.8$)}. 
This IMF has been derived for the Milky Way and represents an intermediate choice between 
the top-heavy and the bottom-heavy IMFs. We also note that this ``Chabrier IMF'' is essentially 
identical to the ``Canonical IMF'' recently summarised by \citet{weidner11}. While we argue that this IMF is the natural choice
on the basis of current evidence, it must still be accepted that we must live with a fundamental uncertainty of $\simeq \times 2$ in stellar masses until 
the uncertainty in the IMF in SMGs is resolved. However, we also note that the derived sSFR  is unaffected
by this IMF uncertainty, making it a particularly useful quantity for comparing the properties of different types of star-forming galaxies.

%-----------------Stellar populations and SFHs-----------
\subsection{Evolutionary models and star-formation histories}
\label{sec:pop}

\begin{figure}
\begin{center}
\includegraphics[width=0.45\textwidth, viewport=5 300 552 550,clip]{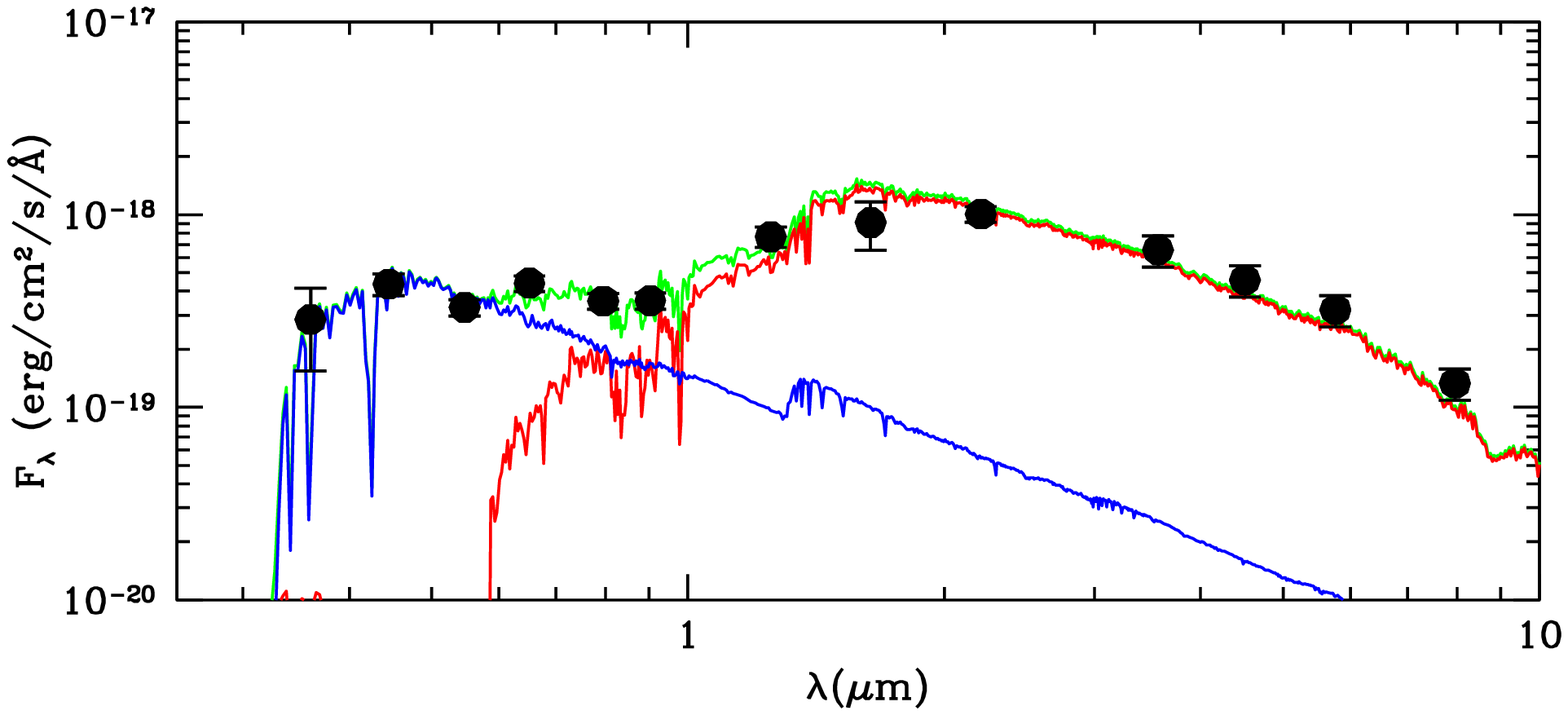}
\end{center}
\caption{ Spectral energy distribution (flux density vs.~observed wavelength) of SMMJ123606.85+621021.4 at the spectroscopic redshift $z=2.509$ fitted using the double-component  star formation history method of \citet[][\citetalias{cirasuolo10}]{cirasuolo10}.  {\it Black circles} denote the data, 
{\it red and blue curves} are the SEDs of the old and young components, respectively, and the {\it green curve} is the sum of these two components. The older component dominates the near-IR light and the derived stellar mass, whereas the young component delivers almost all rest-frame ultraviolet flux density.}
\label{fig:sed}
\end{figure}

\begin{figure*}
\begin{center}
\includegraphics[width=0.9\textwidth,clip]{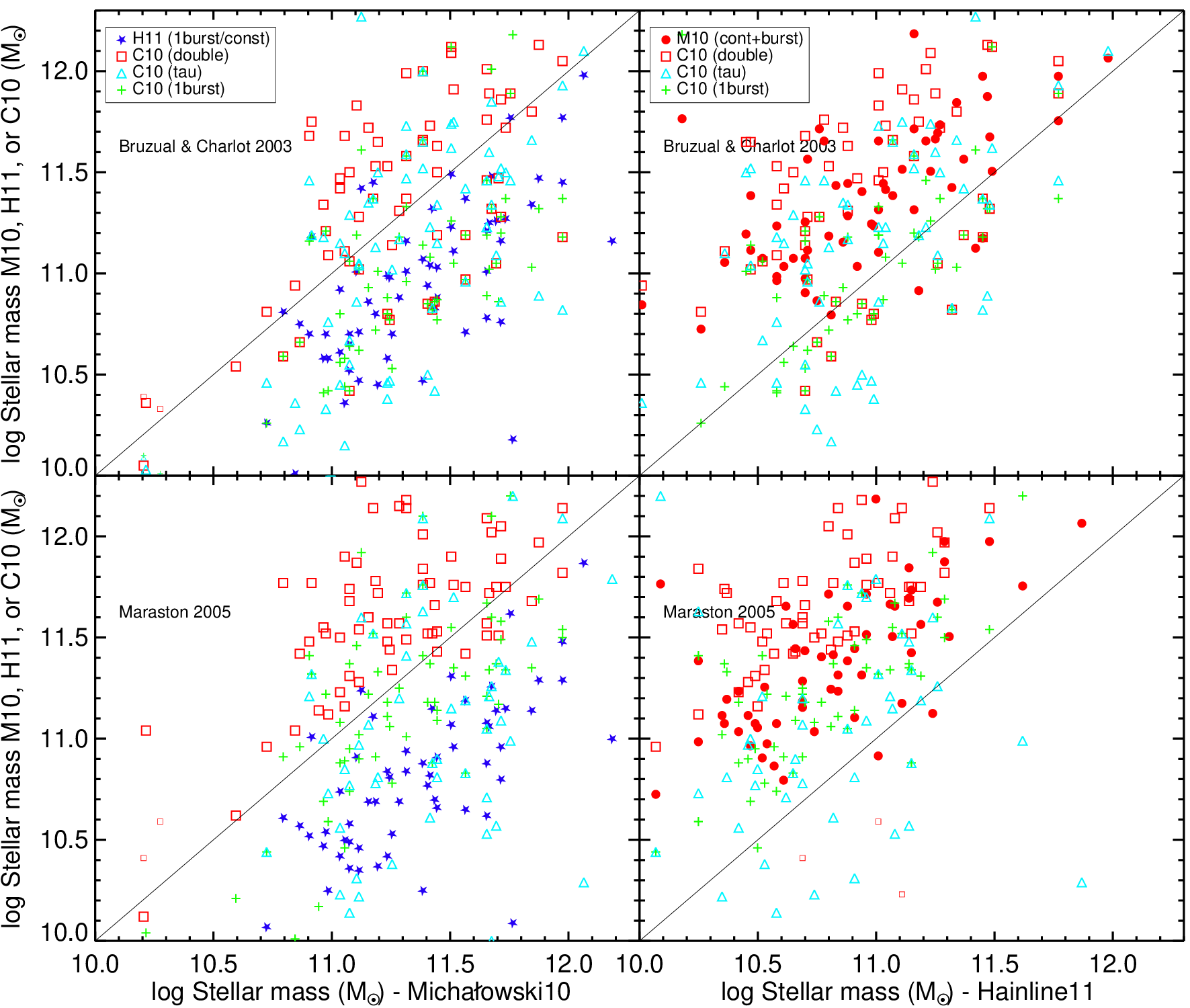}
\end{center}
\caption{ Comparison of the stellar masses of SMGs as derived by {\mich}, {\hain} and in the present work using the stellar population synthesis models of \citet[][{\it top}]{bruzualcharlot03} 
and \citet[][{\it bottom}]{maraston05}. The different symbols indicate the stellar masses derived by {\mich} ({\it red circles}), {\hain} ({\it blue stars}),
and our new estimates based on the  SED modelling method of \citet[][\citetalias{cirasuolo10}]{cirasuolo10} for three alternative 
forms of star formation history: double component model ({\it red squares}), exponentially declining tau model ({\it light blue triangles}) and single burst ({\it green pluses}). 
All the derived masses have been converted to the \citet{chabrier03} IMF. 
Small symbols denote $z<0.5$ SMGs. The {\mich} stellar masses are consistent with our new estimates for the double-component models, 
whereas the {\hain} stellar masses are consistent with single component models, and are on average a factor of $\simeq2.5$ smaller than the double-component estimates (see Tab.~\ref{tab:mcomp}).
}
\label{fig:mcomp}
\end{figure*}

It is well known that the choice of the evolutionary synthesis model used to fit the optical-infrared photometry can affect the derived stellar masses 
of any type of galaxy.
The GRASIL spectral energy distribution (SED) templates \citep{silva98,iglesias07} used by {\mich} make use of the Padova stellar tracks 
\citep{alongi93,bressan93,bertelli94,fagotto94,fagotto94b,fagotto94c,girardi96,girardi00}. 
{ The input metallicity for GRASIL was calculated based on the chemical evolution model in which individual stellar populations are allowed to have metallicity between $0.2$--$5\,Z_\odot$. The mean (median) metallicity derived for SMGs is $0.8\,Z_\odot$ ($0.6\,Z_\odot$).}

 {\hain} used \citet{bruzualcharlot03} 
models (which are also based on the Padova library) and in addition explored the use of the \citet{maraston05} models.  { {\hain} assumed solar metallicity in order to derive a typical mass-to-light ratio.}

The stellar masses derived by {\hain} using \citet{bruzualcharlot03} models are on average $\simeq50\%$ higher than those calculated using the 
\citet{maraston05} models. This average difference is at least in part a consequence of the 
higher contribution of  thermally pulsing asymptotic giant branch (TP-AGB) stars 
in the \citet{maraston05} models, which reduces the derived mass-to-light ratio at rest-frame 
near-infrared wavelengths at stellar population ages $\simeq 1$\,Gyr (a point discussed further below). 

An apparently separate (but, it transpires, in fact connected) cause of differing stellar mass estimates is the chosen parameterisation of the star-formation 
histories (SFHs) of the galaxies in question. As noted by {\hain}, due to their high redshift, and often rather red spectral-energy distributions,
it is very difficult to recover robust SFHs for SMGs from photometric data \citep[cf.][]{dye08,dye10}. 
This leads inevitably to difficulty in constraining galaxy ages and mass-to-light ratios.
 
In an attempt to account for this uncertainty, {\hain} explored the use of two alternative single-component SFHs, adopting either an instantaneous starburst 
or a continuous star-formation history. {\hain} then averaged the two resulting estimates to produce the adopted stellar mass 
for each galaxy. In contrast, {\mich} assumed a two-component SFH consisting of a smooth Schmidt-type law {(SFR$\mbox{}\propto M_{\rm gas}$)} with a recent 
starburst superimposed { commencing $50$\,Myr before the observed galaxy SED was emitted, and with a mass in the range 0--9\% of the infalling gas mass}. 

The assumption of a multi-component SFH generally naturally leads to higher mass-to-light ratios and, in turn, higher stellar masses than the 
use of a single-component model. This is because, while the starburst component can account for the ultraviolet (UV) emission (and the infrared emission 
from the UV-heated dust), the second (older) component is then free to contribute more stars with higher mass-to-light ratios 
(see Fig.~\ref{fig:sed} and also Schael et al. in preparation).
By contrast, the use of a single instantaneous starburst model limits the age of the {\it entire} stellar population to the young age of the 
starburst required to match the UV emission (and thus the true stellar masses are inevitably under-estimated), while in the continuous star-formation model the 
current SFR is set by the UV emission and the total age is then limited in order not to overshoot the optical and the near-IR part of the spectrum 
(assuming the galaxy has always formed stars at the same high rate).

In order to further clarify the impact of the choice of stellar population synthesis models and star-formation histories on the derived stellar masses we 
have revisited the SED fitting of the \citet{chapman05} SMG sample. Specifically, we have used the method presented in \citet{cirasuolo07,cirasuolo10},
based on the {\sc Hyperz} package
 utilising both \citet{bruzualcharlot03}  and \citet{maraston05} models and assuming three different forms of SFH: 
an instantaneous burst of star-formation, an exponentially declining star-formation rate (so-called tau models), and a double-component model composed of two instantaneous
bursts with different ages and (independent) dust attenuations { (an example is shown on Fig.~\ref{fig:sed} { and all estimates are listed in Tab.~\ref{tab:mass}})}. In the double-component model the age of the young component was varied between 
$50$\,Myr and $1.5$\,Gyr, and the old component was allowed to contribute $0$--$100$\% of the near-IR emission while its age was varied over the range $1$--$6$\,Gyr. { Solar metallicity was assumed. 
  These models were run using the Chabrier IMF.

The results of this modelling are illustrated in Fig.~\ref{fig:mcomp} and summarised in Table~\ref{tab:mcomp}, where the mean and median values of $M_*$ and sSFR { (as well as ages and contributions of the older population)} for the SMGs 
are tabulated for the full range of different stellar population synthesis models and star-formation histories explored in {\mich}, {\hain} and the present work 
(it is not possible to distinguish the models based on the goodness of fit).

What is immediately evident from this table is that the results are strongly dependent on the assumed parameterisation 
of the star-formation history { \citep[see also][]{bussmann12}}, as is the impact of assuming different population synthesis models. In particular, it is noteworthy that 
the \citet{maraston05} models only result in significantly lower stellar masses (and hence substantially higher sSFRs) than the \citet{bruzualcharlot03} models 
when a single-component exponentially-declining tau-model SFH is assumed. We investigated the origin of this effect, and found that, when allowed 
only single-component tau models, the mean preferred age of the \citet{maraston05} model fits is $\simeq 0.8$\,Gyr { (see Tab.~\ref{tab:mcomp})}, which corresponds closely to the time 
at which the enhanced TP-AGB star contribution in the \citet{maraston05} models has maximum impact. 
{ This result provides further evidence that the combination of tau SFHs and  the \citet{maraston05} models is not appropriate for SMGs.}

Reassuringly, however, there is no significant difference between the masses derived with these two alternative models for a single burst SFH. 
This is because, in order to reproduce the star-forming nature of SMGs, the single burst fitting typically results in much younger ages of $\simeq0.2$\,Gyr, 
corresponding to the time when the TP-AGB contribution is small and does not have a significant influence on the derived stellar masses. 

As expected (see above discussion of the {\mich} modelling approach), the stellar masses obtained using the double component SFH are systematically larger.
 Especially, the \citet{maraston05} models with the double SFH yield stellar masses higher by $\simeq1$ dex than these derived assuming the tau SFHs with these models.

In summary, the derived stellar masses depend primarily on the assumed parameterisation of the SFH, and the different evolutionary synthesis models only 
yield significantly different results if the single-component exponentially-declining SFH is adopted (due primarily to the extreme behaviour of the 
\citealt{maraston05} models with this adopted parameterization of SFH). Since, as we argue below, a model of a rapidly starforming galaxy in which the SFR is assumed 
to have been even higher at {\it all} previous epochs is obviously physically unrealistic for SMGs, the preferred choice of evolutionary synthesis model is actually of less importance 
in deriving the stellar masses of SMGs than might have been supposed. Nevertheless, in the next section, we do, for completeness, review the arguments 
for and against the choice of the \citet{maraston05} or \citet{bruzualcharlot03} models. 

Of more significance is the fact that, as can be seen from Table~\ref{tab:mcomp}, 
our new two-component models using the 
 \citet{bruzualcharlot03} models 
 yield a value very similar to that derived by {\mich} using a {\it different} form of 
two-component model, and a different evolutionary synthesis model (albeit based on the same stellar tracks as the \citealt{bruzualcharlot03} models).
Thus, consistent average stellar masses are produced by models which allow recent and past SFR to be decoupled, 
regardless of whether the earlier star-formation is parametrised as a single burst or a more extended period of star formation. This consistency
is both reassuring and helpful given that, as we argue further below, a SFH history allowing potentially decoupled recent and past star-formation activity is 
rather obviously appropriate for objects selected on the basis of recent very violent star formation within an already metal enriched environment.
}

%-----------------SSP-----------
\subsection{Choice of evolutionary synthesis model}
{

Recent years have seen some vigorous debate over whether the stronger TP-AGB contribution introduced into evolutionary synthesis models 
by \citet{maraston05} is justified by the data. 
 
Some workers have argued for the superiority of the \citet{maraston05} models based on derived galaxy stellar masses 
\citep{vanderwel06,vanderwel06b}, on derived photometric redshifts \citep{maraston06}, on average broad-band SEDs \citep{maraston06,cimatti08}, 
on optical and near-infrared colours \citep{macarthur10}, and by comparison with semi-analytical models 
\citep{henriques11}. Moreover, \citet{zibetti09} have claimed that stellar mass determinations based on $i$- and $H$-band 
luminosities are more consistent with each other when long-lived TP-AGB stars are introduced.

However \citet{eminian08} showed that \citet{maraston05} models predict excessively red optical colours, and 
recently \citet{kriek10} have found that the average spectral energy distribution of post-starburst galaxies at $0.7 \lesssim z \lesssim 2.0$ does not 
appear to display the excess near-infrared flux predicted from the TP-AGB contribution in the \citet{maraston05} models. Indeed, they demonstrate that 
the \citet{bruzualcharlot03} model provides a much better description of the the data, even at the age when the TP-AGB contribution should be near maximum. 
{ In addition, \citet[][ their Fig.~2 and 16]{conroy10} have found that the \citet{maraston05} models predict excessively red colours for post-starburst galaxies 
and for star clusters in the Magellanic Clouds. 
{  Finally, \citet{zibetti12iau} did not find any strong carbon feature of TP-AGB stars in the near-IR spectra of post-starburst galaxies.}

The work presented here cannot be used to definitively resolve this issue, but we note that the extreme outlying nature of the tau-model \citet{maraston05} results shown in Table~\ref{tab:mcomp}
arguably provides some further evidence against the plausibility and appropriateness of the strong TP-AGB contribution adopted in these models { for tau SFHs}.
{  Moreover, we note that the \citet{maraston05} models with the double SFH yield the highest median stellar masses for SMGs.}

We also note that, confining the comparison to the use of the \citet{chabrier03} IMF and the \citet{bruzualcharlot03} models, the median stellar mass of 
SMGs derived by {\mich} ($M_*=2\times10^{11}\,{\rm M_\odot}$) is in fact only a factor of $\simeq 2.5$ higher than that reported by {\hain} ($M_*=7.6\times10^{10}\,{\rm M_\odot}$).
}

%-----------------Star formation history-----------
\subsection{Choice of star-formation history}

The key issue for stellar mass determination, therefore, appears to be the choice of how to parameterize the SFH of SMGs. Reassuringly, however, the results 
do not appear to be very sensitive to precisely how this is parameterized, provided the SFH allows past and ``present'' star-formation activity to be,
at least in principle, disconnected. This can be seen from the fact that the % 3 
different two-component models {  based on the Padova tracks} listed in Table~\ref{tab:mcomp} yield average stellar masses 
in the range $ M_* = 2$--$3 \times 10^{11}\,{\rm  M_{\odot}}$. 
 
It only therefore remains to consider if 
these two-component models should be preferred to single-component burst or tau models. In fact, at least for SMGs, this is a relatively 
straightforward argument, since both the single-burst and tau models both enforce the condition that the star-formation rates in SMGs were always 
higher at epochs earlier than that at which the object has been selected due to its extreme star-formation rate.
{ Indeed this problem has already been discussed by several authors including \citet{maraston10}. They concluded that, for high-redshift galaxies, 
the assumption of an 
exponentially decreasing SFH (as well as constant SFH to some extent) leads to unrealistically low ages and under-estimated stellar masses, because the SED fit is completely dominated by young stars. 
In fact \citet{maraston10} found an exponentially {\it increasing} SFH could provide a much better description of data. 
Such SFHs are conceptually similar to those used in {\mich}, with the SFR being highest close to the epoch at which the 
observed galaxy SED was emitted, and indeed there is 
growing evidence in favour of rising SFHs in the wider population of galaxies studied at high redshift \citep[e.g.][]{finlator11, papovich11}
}
{
The lack of
either two-component models or rising SFHs 
in the analysis of  {\hain} \citep[or indeed][]{wardlow11} 
can therefore explain all of the remaining discrepancy in the average stellar masses reported by  {\mich} and {\hain} since, as can be seen
from Table~\ref{tab:mcomp}, the use of a double component SFH results in stellar masses of SMGs being $\simeq 3$ times larger than that deduced using the single burst model and $\simeq 2$ times larger than that derived 
using the tau models. This is also illustrated in Figure~\ref{fig:mcomp}, where the masses derived assuming 
single-component SFHs (light blue triangles and green plusses) are systematically lower than those derived assuming the more realistic double-component SFHs (red squares).
This conclusion is consistent with the findings of Schael et al. (in preparation)\footnote{See also \citet{schael09phd}.} who found that the stellar masses of the 
SHADES SMGs \citep{coppin06} derived using a two-component SFH are on average a factor of $\simeq2-3$ higher than when 
derived using a single-component SFH.

}

\begin{table*}
\caption{Stellar, gas and dynamical masses of SMGs (in units of $10^{11}\,{\rm M}_\odot$) \label{tab:mdyn}   }
\centering
\begin{tabular}{lcccccc}
\hline\hline
SMG & $M_{*\rm (M10)}$\tablefootmark{a} & $M_{*\rm (C10)}$\tablefootmark{b} & $M_{*\rm (H11)}$\tablefootmark{c} & $M_{\rm gas}$ & $M_{\rm dyn}^{\rm high CO}$ \tablefootmark{d} & $M_{\rm dyn}^{CO(1-0)}$ \tablefootmark{e} \\
\hline
SMMJ123549.44+621536.8 & $2.66^{+2.65}_{-1.33}$ & $0.66^{+0.66}_{-0.33}$ & 
$2.09^{+0.60}_{-0.47}$ & $2.07\pm1.73$\tablefootmark{f} & 
$0.96\pm0.44$\tablefootmark{g} & $2.30\pm0.40$\tablefootmark{f}\\
SMMJ123618.33+621550.5 & $2.78^{+2.77}_{-1.39}$ & $4.27^{+4.25}_{-2.13}$ & 
$0.76^{+0.09}_{-0.08}$ & $0.35\pm0.11$\tablefootmark{g} & 
$2.18\pm0.18$\tablefootmark{g} & $4.36\pm0.36$\tablefootmark{i}\\
SMMJ123634.51+621241.0 & $2.54^{+2.53}_{-1.27}$ & $0.71^{+0.70}_{-0.35}$ & 
$0.87^{+0.13}_{-0.11}$ & $0.25\pm0.08$\tablefootmark{g} & 
$1.12\pm0.25$\tablefootmark{g} & $2.24\pm0.51$\tablefootmark{i}\\
SMMJ123707.21+621408.1 & $4.95^{+4.93}_{-2.47}$ & $1.12^{+1.12}_{-0.56}$ & 
$1.82^{+0.32}_{-0.27}$ & $1.69\pm1.41$\tablefootmark{f} & 
$2.78\pm1.05$\tablefootmark{g} & $2.20\pm0.60$\tablefootmark{f}\\
SMMJ131201.17+424208.1 & $1.27^{+1.27}_{-0.63}$ & $6.76^{+6.73}_{-3.37}$ & 
$1.02^{+0.33}_{-0.25}$ & $1.34\pm0.50$\tablefootmark{g} & 
$9.50\pm2.38$\tablefootmark{g} & $19.01\pm4.76$\tablefootmark{i}\\
SMMJ163650.43+405734.5 & $2.06^{+2.05}_{-1.03}$ & $3.80^{+3.78}_{-1.90}$ & 
$1.45^{+0.42}_{-0.32}$ & $2.56\pm2.14$\tablefootmark{f} & 
$3.50\pm1.79$\tablefootmark{g} & $4.10\pm0.80$\tablefootmark{f}\\
SMMJ163658.19+410523.8 & $3.27^{+3.26}_{-1.63}$ & $8.13^{+8.09}_{-4.05}$ & 
$1.29^{+0.33}_{-0.26}$ & $2.95\pm2.46$\tablefootmark{f} & 
$1.40\pm0.88$\tablefootmark{g} & $5.80\pm0.40$\tablefootmark{f}\\
SMMJ163706.51+405313.8 & $4.52^{+4.49}_{-2.25}$ & $10.23^{+10.18}_{-5.10}$ & 
$1.62^{+0.24}_{-0.21}$ & $0.24\pm0.07$\tablefootmark{h} & 
$3.40$\tablefootmark{h} & $6.80$\tablefootmark{i}\\
SMMJ221735.15+001537.2 & $1.30^{+1.30}_{-0.65}$ & $1.91^{+1.90}_{-0.95}$ & 
$0.51^{+0.06}_{-0.06}$ & $0.30\pm0.07$\tablefootmark{h} & 
$2.80$\tablefootmark{h} & $5.60$\tablefootmark{i}\\
\hline
\end{tabular}
\tablefoot{ 
\tablefoottext{a}{\citet{michalowski10smg} with \citet{chabrier03} IMF.}
\tablefoottext{b}{From this work using the method of {\ciras} for \citet{bruzualcharlot03} models and double SFH.}
\tablefoottext{c}{\citet{hainline11} for \citet{bruzualcharlot03} models.}
\tablefoottext{d}{Dynamical mass from high exitation CO line.}
\tablefoottext{e}{Dynamical mass from CO(1-0) line.}
\tablefoottext{f}{\citet{ivison11}.}
\tablefoottext{g}{\citet{engel10}.}
\tablefoottext{h}{\citet{greve05}.}
\tablefoottext{i}{Obtained by multiplying the  $M_{\rm dyn}^{\rm high CO}$ by a factor of $2$, the average correction from high excitation lines to CO(1-0) derived for four SMGs with CO(1-0) data.}
Mass errors for {\mich} and {\ciras} are simply a factor of $2$.
}
\end{table*}

\subsection{Other model-fitting issues}

%-----------------Wavelength range-----------
\subsubsection{Wavelength range}
\label{sec:lam}

{ The wavelength range used in the SED model fitting could potentially affect the resulting derived stellar masses. 
For completeness we therefore tested whether the difference in the SMG stellar mass estimates derived by {\mich} and {\hain} 
is not, at least in part, due to the fact that {\mich} fitted the entire multi-frequency dataset up to radio wavelengths, 
whereas {\hain} confined attention to observed $\lambda \leq 8.0\,\mu$m. To test this we simply repeated the analysis 
performed by {\mich}, but this time restricting the model fitting to photometry at optical, near-infrared and IRAC wavelengths. 
We obtained almost identical results 
with a median stellar mass of $\simeq 1.7\times10^{11}\,{\rm M_{\odot}}$ instead of $\simeq 1.9\times10^{11}\,{\rm M_{\odot}}$, and hence 
conclude that the longer-wavelength data have no significant influence on the derived stellar masses.
This is reassuring and is easily explained since the emission at far-infrared wavelengths depends only on the properties of the dust 
and of the younger stellar population that is heating the dust, and should be essentially unaffected by the older stars 
which dominate the total stellar masses of SMGs ({\mich}).
}

%-----------------AGN-----------
\subsubsection{AGN contamination}
\label{sec:agn}

SMGs are believed to suffer only a minor ($\simeq10$--$30$\%) contamination of their bolometric luminosities by active galactic nuclei 
\citep[AGNs;][{\mich}]{alexander05,alexander08,menendezdelmestre07,menendezdelmestre09,valiante07,pope08,hainline09,murphy09, 
watabe09}. Here we consider whether the presence of AGNs in SMGs can significantly influence the derivation of 
their typical stellar masses.

\citetalias{hainline11} reported that, for $\simeq50$\% of SMGs, non-stellar (AGN) power-law features contribute $<10$\% in the rest-frame $H$-band, 
whereas, for $\simeq10$\% of SMGs, the contribution is $>50$\%.  
However, even for those objects which do appear to have a significant AGN contribution at observed IRAC wavelengths, it is not obvious that the stellar masses 
based on our full multi-wavelength fitting will be significantly distorted, especially given the typically large ($\simeq 20\%$) photometric errors generally
assigned to IRAC data in the SED fitting process.
{ 
In fact we find that the stellar masses reported by {\hain} are consistent with our estimates for single component models 
(Fig.~\ref{fig:mcomp} and Tab.~\ref{tab:mcomp}), without making any corrections for possible AGN contributions.
We therefore do not find any evidence to support the idea that the lower stellar masses reported by {\hain} are 
due to the subtraction of an AGN component. Rather, as described above, it seems clear that 
the lower stellar masses deduced by {\hain}  arise, primarily, from the adoption of a single-component SFH.
}

We also note that the presence of any power-law features in the IRAC data should not have strongly affected the mass 
estimates of \citetalias{michalowski10smg}, simply because the 
SED fits were not forced to reproduce the power-law features and, in fact, the best-fitting stellar models tend to 
under-estimate the longer wavelength IRAC data when the power-law component appears to be 
present (see Fig.~A1 of \citetalias{michalowski10smg}). Moreover, we do not find any correlation between the power-law fraction 
given by \citetalias{hainline11} and the stellar masses or $M_*/L_K$ derived by \citetalias{michalowski10smg}, 
which would be expected if the power-law features were biasing the {\mich} stellar mass estimates high.

%-----------------Lensing-----------
\subsubsection{Lensing}
\label{sec:lens}

If SMGs are lensed then the stellar masses derived by {\mich} and {\hain} should be corrected down.
At least some of the brightest SMGs  are thought to be lensed 
\citep{ivison98,downes03,dunlop04,kneib04,motohara05,wilson08b,bercianoalba10,knudsen10,negrello10,swinbank10,vieira10,tamura10,hezaveh11,ikarashi11}. 
However, \citet[][Fig.~2]{lima10} found that the number counts of SMGs with $S_{850}<20$ mJy do not require any lensing correction. 
Moreover, the modelling of \citet[][Fig.~6]{paciga09} showed that only $\lesssim10$\% of SMGs with $S_{850}<15$ mJy at $z<3$ are magnified by a factor of $>2$. 
Hence, the lensing correction for SMGs in the $850\,\mu$m flux-density range considered here is likely to be small on average.

%-------------------------------------------------------------------------------------------
%----------------------------------CONSISTENCY CHECKS----------------------------------
%-------------------------------------------------------------------------------------------
\section{Consistency tests for stellar masses}
\label{sec:check}

%-----------------Mgas Mdyn-----------
\subsection{Dynamical masses}
\label{sec:mdyn}

\begin{figure}
\begin{center}
\includegraphics[width=0.5\textwidth,clip]{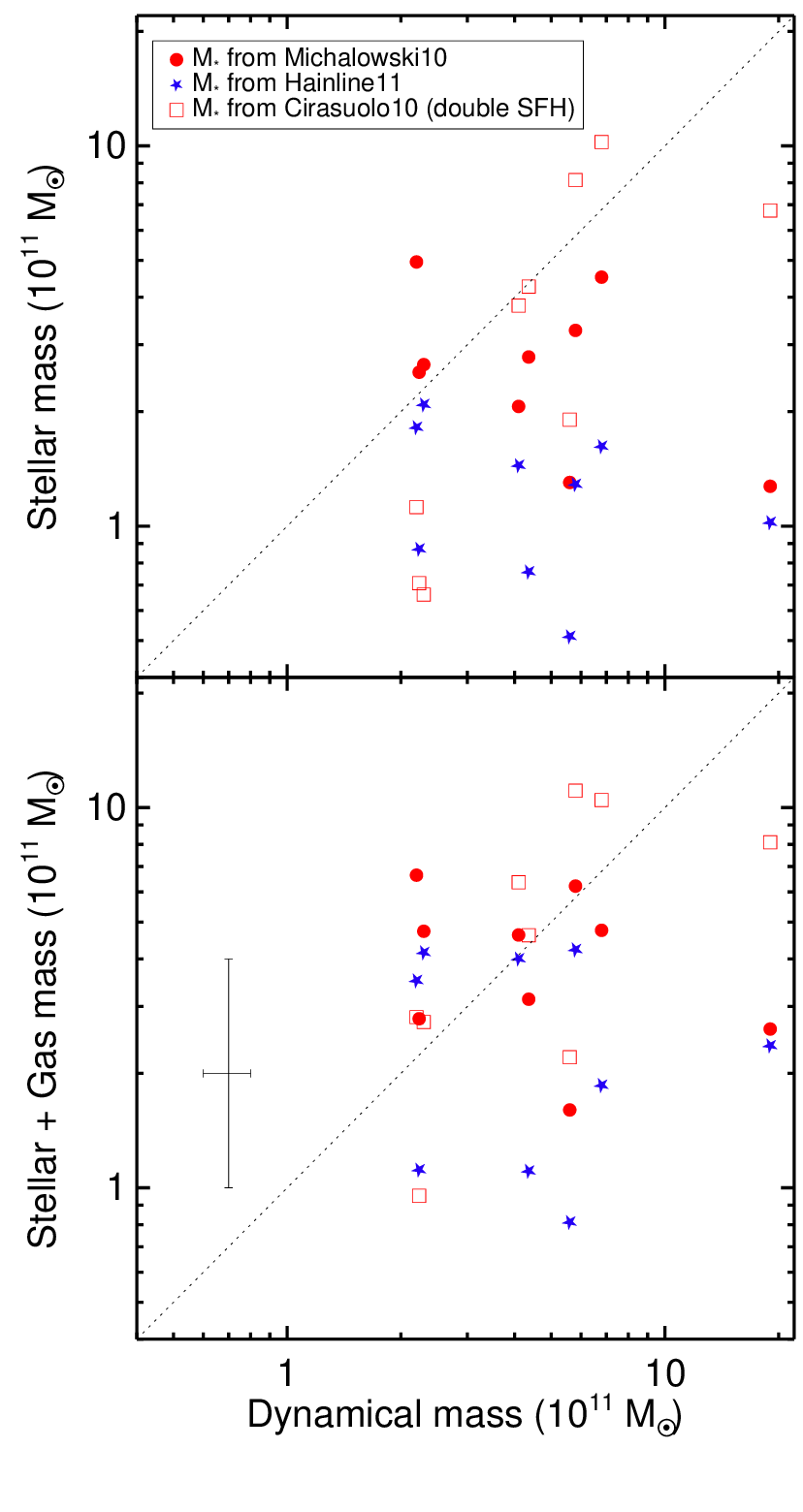}
\end{center}
\caption{Stellar mass ($M_*$) versus  dynamical mass ($M_{\rm dyn}$) estimated from CO(1-0) ({\it upper panel}), and stellar+$H_2$ gas mass ($M_*+M_{\rm gas}$)
versus $M_{\rm dyn}$ ({\it lower panel}) for the 9 SMGs for which the necessary results have been compiled in Table~\ref{tab:mdyn}. 
The data points derived from the stellar mass estimates of {\hain} are indicated by {\it blue stars}, those based on the stellar mass estimates of {\mich} are indicated 
by {\it red circles}, while the new double-component {\ciras} values are indicated by {\it red squares}.  
In principle, only the area below the {\it dotted line} represents a consistent set of galaxy properties with $M_{\rm gas}+M_*\leq M_{\rm dyn}$, but because of 
the typical scatter (indicated by the error bars in the lower panel) and unknown 
inclinations, $\simeq 2-3$ out of the 9 
individual SMGs are in fact expected to exceed this 
apparent threshold (see Section~\ref{sec:mdyn} for details).
We see that, contrary to the conclusion of  \citet{engel10}, there is no evidence that the stellar masses derived by 
\citetalias{michalowski10smg} using the Chabrier IMF  are in conflict with the dynamical mass estimates, even allowing for estimates of the 
additional gas mass. Indeed, the average values of these 
ratios favour the {\mich} stellar mass estimates as the most 
consistent with the available data (see Section~\ref{sec:mdyn}).}
\label{fig:mgas}
\end{figure}

The estimates of the dynamical and $H_2$ gas masses from CO lines detected for SMGs provide a consistency check on the stellar mass estimates. 
Specifically, assuming that the extent of the CO-emitting gas is the same as the extent of the stellar component, the sum of the gas and stellar masses 
has to be less than the dynamical mass. Using this approach \citet{engel10} derived a mean ratio  $(M_*+M_{\rm gas})/M_{\rm dyn}\simeq2.1$ for seven SMGs using the stellar masses of {\mich} (corrected to the \citealt{chabrier03} IMF) and concluded that the stellar masses derived by \citetalias{michalowski10smg}   are inconsistent with their gas and dynamical masses. As explained below, we do not find 
that this is the case.

There is, of course, some uncertainty and controversy over the calculation of dynamical and molecular gas masses from CO line observations \citep[e.g.][]{genzel03,ivison11}. 
Perhaps of most relevance for the present discussion is the fact that \citet{engel10} used high excitation lines (CO J=7-6, 6-5, 4-3 and 3-2), 
which trace denser and hotter gas, very likely 
confined to the more central regions of a galaxy. Therefore, the derived dynamical masses could easily refer to regions smaller than the extent of the stellar 
component, making it difficult to compare them directly with the stellar masses. Indeed, based on low excitation CO(1-0) lines \citet{ivison11} found  
dynamical masses higher by a factor $\simeq2.4$--$4.1$ for two out of three SMGs, for which  \citet{engel10} claimed a significant discrepancy 
between stellar and dynamical mass (HDF 76 and N2 850.2; the third one was not observed by \citeauthor{ivison11}). 
{ Moreover, recently \citet{riechers11} found CO(1-0) sizes  $\simeq60$--$90$\% larger than that reported by \citet{engel10} for two additional SMGs.
}

{ Interestingly, the mean CO half-light radius obtained by \citet{engel10} of $2.4$\,kpc is indeed $15$--$40$\% smaller than the typical half-light 
radius of the near-infrared emission dominated by the stellar component \citep[$2.8$--$3.4$ kpc;][]{swinbank10b,mosleh11,targett11}. 
Moreover, one out of three SMGs for which 
\citet{engel10} claimed inconsistency with the {\mich} stellar mass (N2 850.2) was observed by \citet{swinbank10b} who measured the stellar 
extent to be $\simeq3$ times larger than the extent of the 
\mbox{CO(7-6)} found by \citet{engel10}. This reinforces the likelihood that the dynamical masses of the problematic SMGs 
have been under-estimated, as the stellar component is more extended than the high-excitation gas.
}

This argument is supported by hydrodynamical simulations of SMGs  \citep{narayanan09,narayanan10,hayward11b,hayward11}, where only $\simeq30$\% of the 
stellar mass is distributed over the inner radius of $2.5$\,kpc, roughly the size over which \citet{engel10} calculated dynamical masses. This correction 
to the dynamical masses would be enough to remove the inconsistency with the stellar masses of {\mich}, even though some simulations 
predict that the usual conversions result in the dynamical masses overpredicted by a factor of  $\simeq1.5$--$1.9$ \citep[][Fig.~10 and Eq.~6]{narayanan09}.

In an attempt to overcome these systematics, and facilitate a fairer comparison of dynamical and stellar masses, we 
focus here on the newly available CO (1-0) dynamical masses which, as described above, appear to better trace the full extent of 
the SMGs, yielding somewhat larger sizes and masses. Stellar, gas and dynamical masses are available for 9 of the galaxies in our 
sample, including two CO detections made by \citet{greve05}. These data\footnote{Note that \citet{engel10} used the stellar masses reported 
initially in \citet{hainline08phd} whereas we quote the corrected values from \citetalias{hainline11}.} are presented in Table~\ref{tab:mdyn}. 

In the final column of this table we give the CO (1-0) dynamical mass as our best 
estimate of the dynamical mass of each galaxy. Where available (4 SMGs) this number has been taken directly from the CO (1-0) 
observations performed by \citet{ivison11}, while for the remaining objects this number has been inferred from the higher-order 
CO-line dynamical masses by multiplying them by the average correction factor for the aforementioned 
4 SMGs for which the dynamical masses of \citet{ivison11} and  \citet{engel10} can be directly compared.

In Fig.~\ref{fig:mgas} we then plot, for these 9 SMGs,  $M_*$ and also $M_{\rm gas}+M_*$ as a function of $M_{\rm dyn}$, with the dynamical masses taken from the 
final column in Table~\ref{tab:mdyn}, and the stellar mass estimates from {\hain}, {\mich} and the present work (double-component {\ciras} model) indicated 
by the different symbols.

From Fig.~\ref{fig:mgas} it can be seen that, within the errors, there is no evidence that even the higher stellar mass determinations for any object exceed the estimated 
dynamical mass. At first sight it might seem that there could be more of a problem (for all mass determinations) for  $M_*+M_{\rm gas}$, 
but this is not really the case, both because of uncertainty in $M_{\rm gas}$, and because of the potential effects of varying inclination angle on $M_{\rm dyn}$.
Specifically, while the unknown inclination ($i$) of gas discs should not influence the average $(M_{\rm gas}+M_*)/M_{\rm dyn}$ ratio, 
if the real inclination of an {\it individual} galaxy is lower than 
the assumed average, then its derived $M_{\rm dyn}$ is inevitably too low, which can make it {\it appear} to be inconsistent with
the (orientation-independent) derived stellar mass. To quantify this effect we considered 100,000 galaxies with real $(M_{\rm gas}+M_*)/M_{\rm dyn}=1$ (or $0.5$) 
with inclinations distributed uniformly in three dimensions, i.e. with a mean $\langle\sin^2 i\rangle=2/3$. We found that the adoption 
of this mean value for all galaxies results in $\simeq18\%$ ($9$\%) of them {\it appearing} to exceed the allowed $(M_{\rm gas}+M_*)/M_{\rm dyn}=1$ 
by a factor of $>2$ and $\simeq12$\% ($7$\%) by a factor of $>3$. Hence, in a sample of $\simeq10$ galaxies with 
random unknown inclinations,  we would expect $1$--$2$ of them to significantly exceed the allowed ratio, simply because of
the unknown inclination of any molecular disc in an individual object.

In summary, one should expect, given the effects of random inclination and uncertainty in the measurement of $M_*$, $M_{\rm gas}$ and $M_{\rm dyn}$, that a few galaxies should
lie above the one-to-one relation indicated by the dotted line in Fig.~\ref{fig:mgas}, and it is the mean (or median) value of the mass ratios which need to be considered when 
reviewing the viability of the stellar masses. Working from the numbers given in Table~\ref{tab:mdyn}, for the 
ratio $M_*/M_{\rm dyn}$ the mean (median) values are 0.36 (0.24) for the {\hain} stellar masses,
0.80 (0.64) for {\mich}, and 0.94 (0.78) for the new double-component {\ciras} values. Thus, adoption of the lower {\hain} stellar masses would imply 
that only $\simeq 1/3$ of the dynamical mass of a typical SMG can be accounted for by $M_*$. In contrast, the {\mich} and {\ciras} results 
would imply that $M_*$ for a typical SMG is $\simeq 3/4 \times M_{\rm dyn}$. If the values of $M_{\rm gas}$ in Table~\ref{tab:mdyn} are taken at face value
then this latter scenario would appear to be more consistent with the data, as 
for the ratio 
$(M_*+M_{\rm gas})/M_{\rm dyn}$ the mean (median) values are 0.71 (0.50) for the {\hain} stellar masses,
1.15 (1.07) for {\mich}, and 1.09 (1.19) for the new double-component {\ciras} values. This then provides further evidence that the stellar mass estimates of {\hain} are a little too low, the new {\ciras} double-component values are perhaps a little high, and the {\mich} results appear to provide the best description of the data.

We conclude that the latest estimates of the dynamical masses of SMGs are in very good accord with the stellar masses of SMGs estimated 
by {\mich} with the \citet{chabrier03} IMF.

%-----------------number density-----------
\subsection{Number density of massive galaxies}
\label{sec:n}

Another consistency check on the plausibility of the stellar-mass estimates is to assess how the number density of SMGs compares with the 
number density of galaxies in general in the same implied stellar-mass range, at both lower and comparable redshifts.

\citet{drory04} reported the cumulative number density of galaxies $n(M_*> 1.1\times10^{11}\,{\rm M_\odot}) = 3.16$--$0.64\times 10^{-4}$\,Mpc$^{-3}$ at redshifts $z=0.5-1.1$ 
(their Fig.~6 and Table~3, after conversion of their mass limit to the \citealt{chabrier03} IMF). Restricting the analysis to the SMGs at $z>1.4$, and using the comoving 
volumes given by  \citetalias{michalowski10smg} we obtain $n_{\rm SMG}(M_*> 1.1\times10^{11}\,{\rm M_\odot}) = 7.0\times 10^{-6}$\,Mpc$^{-3}$ for 
the $M_*$ values of \citetalias{michalowski10smg} and $3.8\times 10^{-6}$ Mpc$^{-3}$ for the $M_*$ values of \citetalias{hainline11}. Hence, even for the {\mich} stellar mass estimates the number density of SMGs is 
$\simeq 30$ times lower than that of massive galaxies at lower redshifts { and therefore the number of massive SMGs is clearly not inconsistent with the number density of massive galaxies at lower redshifts,
even allowing for a relatively modest and plausible duty cycle (see below).
In addition, from the mass comparisons discussed in the previous subsection, further conversion of the observed molecular gas reservoir into stars, even if extremely efficient, can at most increase the measured {\mich} stellar masses
by $\simeq 30-50$\% (or, equivalently, the {\hain} masses by $\simeq 100$\%), and so this basic conclusion is not seriously affected by potential further star formation. 
}

The redshift range $1.4<z<3.6$ corresponds to a time interval of $2.7$\,Gyr. Given the typical gas reservoirs of $M_{gas} \simeq 1 \times 10^{11}\,{\rm M_{\odot}}$ listed 
in Table~\ref{tab:mdyn} it can be seen that the typical gas exhaustion timescale is $\simeq 200$\,Myr (assuming a typical SFR of $\simeq 500 \,{\rm M_{\odot}\, yr^{-1}}$). If 
one further assumes that a typical SMG is observed mid-way through its final violent star-forming peak, then a reasonable estimate of the {\submm} luminous 
lifetime of an SMG is  $400$\,Myr \citep[see also][]{bouche05}. This implies a duty-cycle number density correction factor of $\simeq 6-7$ for the redshift 
interval in question. If one further factors an  incompleteness correction of a factor of $\simeq 3.5$ as derived by \citetalias{michalowski10smg}, 
then the implied true number density of {\it potential} SMGs in this era is close to, but still less than the best estimate of 
$M_*> 1.1\times10^{11}\,M_\odot$ galaxies at $z\simeq1.1$. Hence, the conclusion that the stellar masses of SMGs are typically 
$M_* \simeq 2 \times 10^{11}\,{\rm M_{\odot}}$ appears to be perfectly consistent with the number density of massive galaxies at $z \simeq 1$.

A similar conclusion can be reached comparing to the number density reported by \citet[][Fig.~11]{pozzetti07} and \citet[][Fig.~16]{kajisawa09}:
$n(M_*> 6\times10^{10}\,{\rm M_\odot}) \simeq 10^{-3}$--$2\times10^{-4}$ Mpc$^{-3}$ at $z=0$--$2$, whereas the corresponding values for SMGs 
are $n_{\rm SMG}(M_*> 6\times10^{10}\,{\rm M_\odot}) = 8.7\times 10^{-6}$ and $5.6\times 10^{-6}$ Mpc$^{-3}$.

Finally, we integrated the mass function at $z=3$--$3.5$ given by \citet{caputi11} for $M_*> 10^{11}\,{\rm M_\odot}$ galaxies, obtaining $7\times10^{-6}$ Mpc$^{-3}$.
This rather narrow redshift interval corresponds to an elapsed time of only 350\,Myr, and so, following the above arguments, the duty cycle correction
should be $\simeq$ unity. It is therefore perhaps unsurprising that this number density is rather similar to the 
number density of SMGs. 

\begin{figure}
\begin{center}
\includegraphics[width=0.5\textwidth,clip]{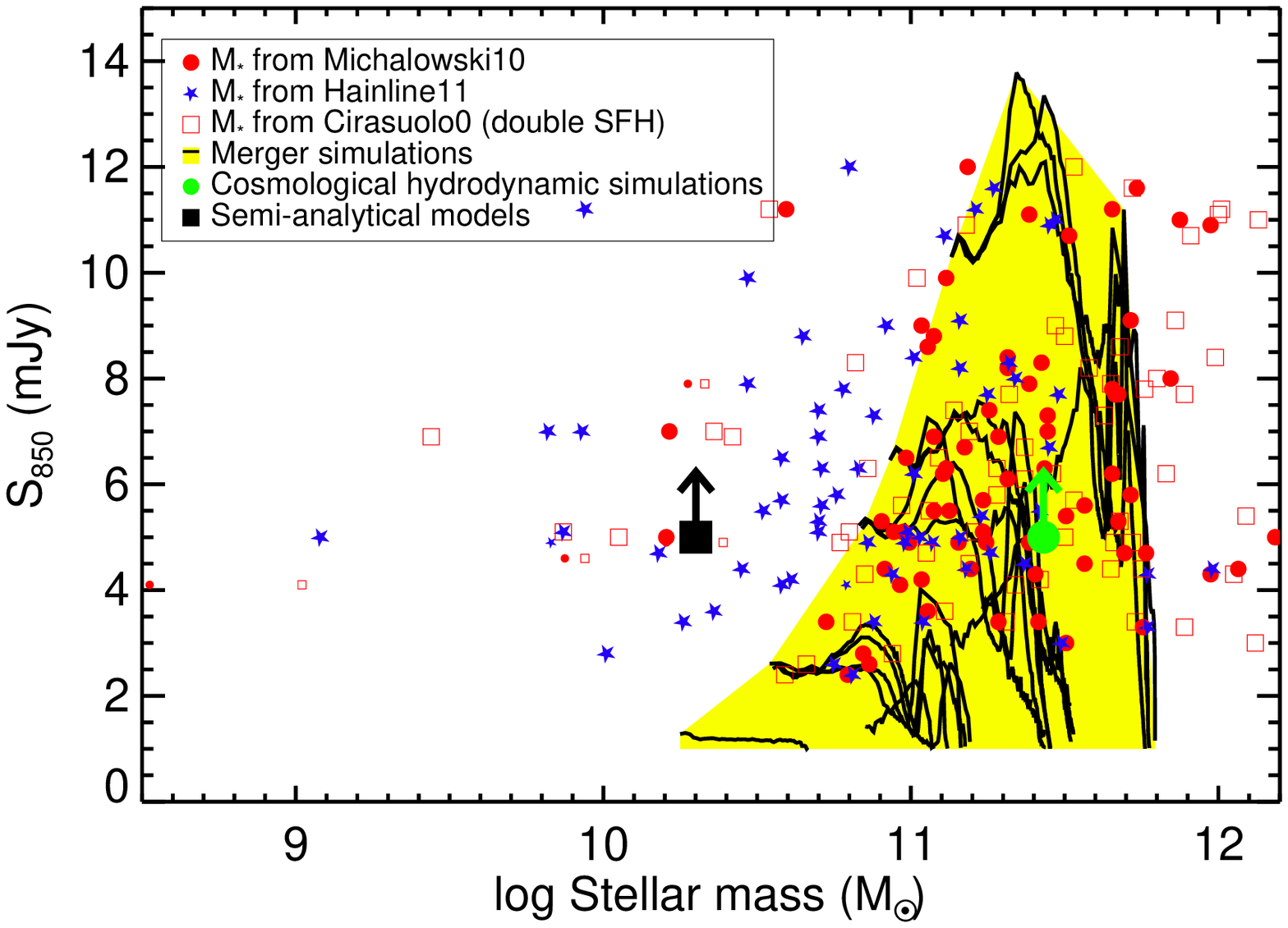}
\end{center}
\caption{Submillimetre flux densities as a function of the stellar masses of SMGs. 
{\it Solid lines} represent the evolutionary tracks from the merger hydrodynamic simulations 
\citep[][]{narayanan09,narayanan10b,narayanan10,hayward11b,hayward11} enclosed in the {\it yellow shaded area}, 
{\it the big square} represents the median stellar mass of $2\times10^{10}\,M_\odot$ for SMGs with $S_{850}>5$ mJy 
in the semi-analytical models \citep{baugh05,swinbank08,gonzalez11,almeida11}, whereas {\it the big circle} corresponds to the median value derived by \citet{dave10} from the cosmological hydrodynamic simulations.
Other symbols  
represent the observational data for SMGs with stellar masses from {\mich} ({\it red circles}), 
{\hain} ({\it blue stars}) and our new estimates based on {\ciras} and double-component SFH ({\it red squares}). Small symbols indicate SMGs at $z<0.5$. According to the hydrodynamic simulations, 
many of the stellar masses derived by {\hain} are too low to result in measurable {\submm} flux, while the stellar masses 
derived by {\mich} are consistent with the hydrodynamic simulation predictions (within 
the observational errors). Neither the {\hain}, {\mich} or {\ciras} stellar masses appear to agree, on average, 
with the median stellar mass predicted by the semi-analytic models. 
}
\label{fig:f850}
\end{figure}

%-----------------Simulation-----------
\subsection{Numerical simulations of SMGs}
\label{sec:sim}

Figure~\ref{fig:f850} shows {\submm} flux density versus stellar mass for a set of idealized
hydrodynamic simulations of merging and isolated disk galaxies, where
3-D radiative transfer has been performed to calculate the {\submm} flux.
The simulations are similar to those described in \citet{narayanan09,narayanan10b,narayanan10}; 
see \citet{hayward11b,hayward11} for a description of the
differences. Note that the plot simply shows tracks for different
simulations. No weighting by abundance has been applied, so only the
range spanned by the simulations, not typical values, can be read from
the plot.

We note that in order to reach $S_{850\,\mu{\rm m}}
\gtrsim 5$\,mJy, these simulations require stellar mass to exceed
$M_* \simeq 1 \times
10^{11}\,{\rm M_{\odot}}$. As can be seen from Fig.~\ref{fig:f850}, the predicted stellar 
mass range is in good agreement with the SMG stellar masses derived by {\mich}. 
We note that, in
the cold infall model of SMGs, comparably high stellar masses are also
preferred \citep[][their Fig.~2]{dave10}.

On the other hand, the semi-analytical models of \citet{baugh05}, \citet{swinbank08}, \citet[][Fig.~3]{gonzalez11} 
and \citet{almeida11} predict significantly lower stellar masses for SMGs,
with a median  $M_* \simeq 2\times10^{10}\,M_\odot$ (indicated by the square on Fig.~\ref{fig:f850}). 
Such a value is lower even than the vast majority of the stellar masses derived by  {\hain},  
confirming the findings of \citet{swinbank08} that the stellar masses of SMGs are generally 
systematically under-predicted by this class of semi-analytic model.

%-------------------------------------------------------------------------------------------
%----------------------------------SSFR----------------------------------
%-------------------------------------------------------------------------------------------
\section{Specific Star-Formation Rates}
\label{sec:SSFR}

 \begin{figure*}
\begin{center}
\includegraphics[width=0.80\textwidth,clip]{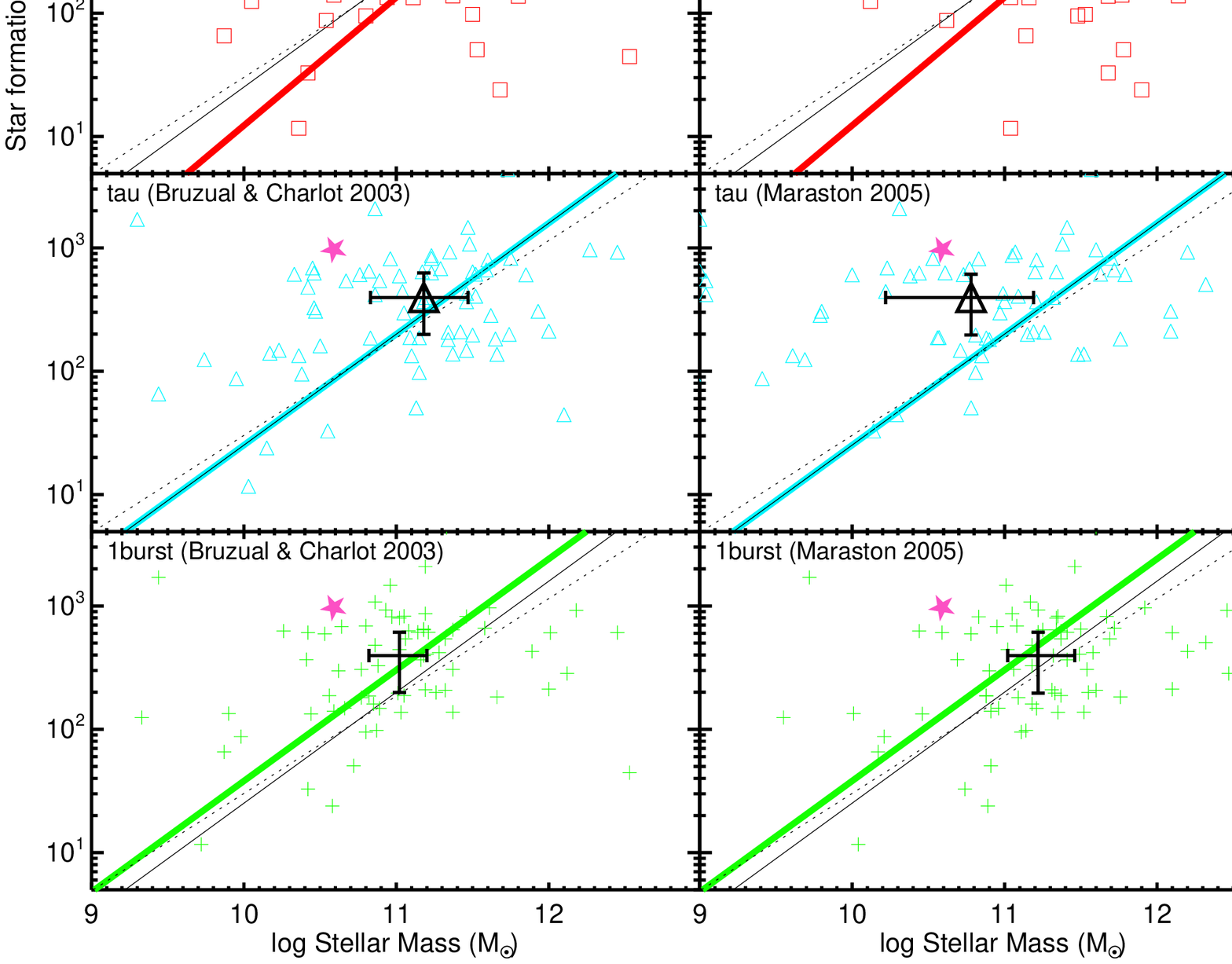}
\end{center}
\caption{ SFRs of SMGs as a function of stellar masses  derived  using the stellar population synthesis models of \citet[][{\it left}]{bruzualcharlot03} and \citet[][{\it right}]{maraston05}. 
The symbols correspond to the stellar masses derived by {\mich} ({\it red circles}), {\hain} ({\it blue stars}) and our new estimates based on the  SED modelling method 
of \citet[][\citetalias{cirasuolo10}]{cirasuolo10} for three alternative forms of star formation history (SFH): double-component model ({\it red squares}), 
exponentially-declining (tau) model ({\it light blue triangles}) and single burst ({\it green pluses}). The median values are shown as 
{\it large black symbols} with $3\sigma$ error bars. The `main-sequence' of star-forming galaxies at $z \simeq 2$ is shown as a 
{\it solid line} \citep{daddi07} and {\it dotted line} \citep{rodighiero11}. { The {\it thick, coloured solid lines} represent the  `main-sequence'  of \citet{daddi07} at $z\simeq2$ corrected for the systematic difference in average stellar mass introduced by the  self-consistent adoption of SFH to model SMGs in each panel (see Section~\ref{sec:SSFR} for details).} All results have been converted to the \citet{chabrier03} IMF. 
Unless the exponentially declining (tau model) SFH with the \citet{maraston05} models %or the single burst SFH with the  \citet{bruzualcharlot03} models 
are assumed, 
the SMGs are completely consistent with the high-mass end of this sequence. This agreement is essentially perfect for the {\mich} and double-component models.
The {\it magenta star} shows the position of an average SMG as plotted by \citet{daddi07}. Clearly, regardless of the assumed  evolutionary stellar model or SFH, this 
supposed typical value of sSFR for SMGs is in fact unrepresentative of real SMGs.
}
\label{fig:sfrm}
\end{figure*}

One important step towards locating SMGs within the general framework of galaxy/star formation at high redshift is to establish whether, on average, they simply represent the most massive extension of 
the normal star-forming galaxy population in the young active Universe at $z > 2$ (as predicted by \citealt{dave10} in cosmological hydrodynamic simulations), or whether they are extreme outliers in terms of specific star-formation rate (sSFR).

To explore this we compare the properties of SMGs with other galaxies at similar redshifts as presented by \citet[][Fig.~17]{daddi07}, 
\citet[][Fig.~2]{kajisawa10} and \citet[][Fig.~1]{rodighiero11}. 
{
For the SMGs we adopt the SFRs derived by {\mich} from the mid-far infrared emission. 
The comparison is shown in Figure~\ref{fig:sfrm}. Here, for the nine different determinations of SMG masses considered in this paper, 
we plot SFR versus $M_*$ for all the individual SMGs in the sample, and compare the results with the location
of the so-called `main-sequence' of star-forming galaxies at $z \simeq 2$ \citep{daddi07,rodighiero11}. 

{
In order to ensure that this  comparison is fair and self-consistent, we have also investigated how the choice of SFH influences the location of this `main-sequence'. We assumed that the SFR estimate is independent of  the adopted SFH (as it is usually derived from bolometric or monochromatic luminosity), and derived the stellar masses of 
%all galaxies in the CANDELS/UDS field 
all galaxies the UDS field in CANDELS \citep{grogin11,koekemoer11}
at $z=1.5$--$3$ and $M_*=10^{9.5}$--$10^{11.6}\,{\rm  M_{\odot}}$ (Cirasuolo et al.~in prep.) using the same SED fitting method outlined in Section \ref{sec:pop} for double-burst, single-burst and tau SFHs. If we adopt the results of the single-burst models for these galaxies, we obtained masses consistently lower by $0.1$ dex relative to those derived using the tau models; on Fig.~\ref{fig:sfrm} (`1burst' panels) this is indicated by the position of the coloured thick line towards lower stellar masses. On the other hand, as expected, the adoption of the double-burst models yields higher stellar masses. However, this shift transpires to be mass-dependent; at $10^{10}\,{\rm  M_{\odot}}$ the double-burst SFH produces stellar masses $0.3$ dex higher than the tau SFH, whereas this shift is only $0.1$ dex at $10^{11.5}\,{\rm  M_{\odot}}$ (resulting in steepening of the thick line in the `double' panels). {  The latter} is a significantly smaller shift than the $0.3$ dex difference found between the double-burst and tau model stellar masses for SMGs (Tab.~\ref{tab:mcomp}).
}

It is clear from this figure 
that, for {almost} all models 
 the SMGs lie essentially exactly on this `main-sequence', within the errors. It is also 
clear that, even for the lower mass determinations, the average deviation from the `main-sequence' is relatively modest. As expected (given Table~\ref{tab:mcomp} and the discussion in 
Section 2.2), the offset is greatest for the exponentially-decreasing tau SFH with the \citet{maraston05} models, but even then it can be seen that the majority 
of the SMGs are still actually consistent with the `main-sequence', and the formal $\simeq 3\sigma$ offset in average sSFR is in fact driven by a subset of extreme outliers.
It can also be seen that none of the various determinations of SMG stellar masses yields a sSFR for a typical SMG as extreme as that claimed 
by \citet{daddi07}. 
{ This discrepancy is likely because their stellar mass estimates are based on dynamical masses from high-excitation CO line observations \citep{tacconi06}, which may bias the  stellar mass estimates low, as we explained in Section~\ref{sec:mdyn}. Moreover, their use of infrared luminosity of $\sim10^{13}\,{\rm  L_{\odot}}$, estimated from a single $850\,\mu$m detection \citep{greve05,tacconi06}, represents an overestimate, as revealed by recent {\it Herschel} studies \citep[][$\sim10^{12.5-12.7}\,{\rm  L_{\odot}}$]{magnelli12,roseboom12} making the SFR estimate of \citet{daddi07} a factor of $2$--$3$ too high.}
 
To quantify the apparent consistency between the properties of SMGs and other star-forming galaxies at comparable epochs we report, in Table~\ref{tab:mcomp}, 
the mean and median sSFRs for the SMGs as derived from the various alternative models (as stated above these values are IMF-independent). 
Again, with the possible exception of the \citet{maraston05} tau-model, all the median values are clearly consistent with the values of a few Gyr$^{-1}$ 
found by \citet{daddi07} and  \citet{kajisawa10}. Hence, it would appear that, while a few SMGs may be caught at unusually spectacular moments, 
in general it seems hard to avoid the conclusion that the SMG population is perfectly consistent with the high-mass end of the general star-forming galaxy population at these redshifts.
Interestingly, it is only really the double-component models which yield mean and median values which are formally consistent with each other, with the other models always producing a 
few extreme outliers with very low stellar masses. Thus, the use of the most physically plausible star-formation histories yields consistent and fairly well 
defined typical mean sSFR values for SMGs of $2 - 3$\,Gyr$^{-1}$, in rather good agreement with the height of the plateau in sSFR  at $z > 2$ discussed by \citet{gonzalez10}.
}

%----------------------------------CONCLUSION----------------------------------
\section{Conclusions}
\label{sec:conclusion}

We have explored the potential origins of systematic differences in the stellar masses derived for SMGs, and have found that the  
difference between the stellar-mass estimates of \citetalias{michalowski10smg} and \citetalias{hainline11} can be explained by 
the combination of different assumed IMFs, stellar population models and types of star formation histories. The available evidence 
supports the use of the Chabrier IMF (with the Salpeter IMF yielding excessive masses for several SMGs).
{ The choice of which stellar synthesis model to adopt turns out to be relatively unimportant compared to the choice of star formation history. The one exception to this is the combination of an exponentially declining  star formation history with the models of \citet{maraston05}, which produces radically different, unrealistically low stellar masses, presumably due to the very strong assumed TP-AGB contribution in these models.}

For the same IMF and evolutionary synthesis model, the remaining difference between the average stellar mass estimates 
of \citetalias{michalowski10smg} and \citetalias{hainline11} can be accounted for by differences in adopted star-formation history. Specifically,
the two-component models of \citetalias{michalowski10smg} yield values $\simeq 3$ times larger than the single-component models 
utilised by \citetalias{hainline11}. We have explored this issue in more detail, by running further single-burst, tau-model and double-component model fits 
using the method of {\ciras} and using both the \citet{bruzualcharlot03} and  \citet{maraston05} models. We find that the models which allow 
past and ``present'' star-formation activity to be disconnected yield very consistent average stellar masses in the range 
$ M_* = 2$--$3 \times 10^{11}\,{\rm  M_{\odot}}$, regardless of precisely how this is parameterized.
We argue, therefore, that the main reason that authors such as \citetalias{hainline11} and \citet{wardlow11} have concluded 
in favour of typical SMG stellar masses $ M_* < 1 \times 10^{11}\,{\rm  M_{\odot}}$ is that they have assumed models in which the 
SFR in SMGs was always as high or higher prior to the epoch at which the object has been selected on the basis of its extreme 
star-formation activity. The more physically realistic two-component models are supported by other evidence for rising star-formation 
histories in star-forming galaxies at $z > 2$.

Considering potential orientation effects, and updated estimates of dynamical masses,
we find that even the higher stellar mass estimates of \citetalias{michalowski10smg} { with the Chabrier IMF} certainly cannot be excluded with current CO line data, and indeed 
yield total masses in good agreement with dynamical mass estimates.
We also argue that SMG stellar masses $ M_* \simeq 2 \times 10^{11}\,{\rm  M_{\odot}}$ appear to be perfectly plausible given current measurements of the evolving mass function of the general galaxy population
selected at near-infrared and mid-infrared wavelengths.

We also show that the mass estimates of {\mich} are in good agreement with 
the predictions of hydrodynamical merger models of SMGs and the cold inflow scenario, whereas those of {\hain} are closer to 
(but still not as low as) the much lower-mass predictions of certain semi-analytic models  of galaxy formation.

Finally, we find that the inferred specific star-formation rates of SMGs are perfectly consistent 
with those derived for other star-forming galaxies 
at comparable redshifts, suggesting that most SMGs simply represent the high-mass extension of 
the `main-sequence' of star-forming galaxies at $z \ge 2$, and in general should not 
be considered as outliers in the stellar mass-SFR relation.

%----------------------------------acknowledgments----------------------------------

\begin{acknowledgements}

We thank Joanna Baradziej and our anonymous referee for help with improving this paper and Georgios Magdis for comments.
MJM and JSD acknowledge the support of the UK Science \& Technology Facilities Council. JSD acknowledges the 
support of the Royal Society via a Wolfson Research Merit award, and the support of the European Research Council 
through an Advanced Grant. 
The Dark Cosmology Centre is funded by the Danish National Research Foundation. 
This research has made use of Tool for OPerations on Catalogues And Tables \citep[TOPCAT;][]{topcat}: \url{www.starlink.ac.uk/topcat/ } and NASA's Astrophysics Data System Bibliographic Services.

\end{acknowledgements}

\appendix
\section{Table with all stellar mass estimates}
\begin{table*}
\scriptsize
\caption{Stellar masses of individual SMGs calculated using various methods, stellar population models and star formation histories. \label{tab:mass}   }
\centering
\begin{tabular}{lcrrrrrrrrr}
\hline\hline
& & \multicolumn{9}{c}{$M_*$ ($\log M_\odot$)} \\ \cline{3-11}
 & & M10\tablefootmark{a} & H11\tablefootmark{b} & H11\tablefootmark{c} & C10\tablefootmark{d} & C10\tablefootmark{e} & C10\tablefootmark{f} & C10\tablefootmark{g} & C10\tablefootmark{h} & C10\tablefootmark{i} \\
 & & Padova & BC03 & M05 & BC03 & BC03 & BC03 & M05 & M05 & M05 \\
 SMG & $z$ & double & single & single & 1burst & tau & 2burst & 1burst & tau & 2burst \\
\hline
SMMJ030226.17+000624.5 & 0.080 & 10.275 & -99.000 & -99.000 & 10.010 & 10.000 & 10.330 & 9.810 & 9.810 & 10.590 \\
SMMJ030227.73+000653.5 & 1.408 & 10.915 & 11.180 & 11.009 & 11.180 & 11.190 & 11.750 & 11.320 & 11.320 & 11.770 \\
SMMJ030231.81+001031.3 & 1.316 & 10.205 & 9.080 & 9.209 & 9.330 & 9.740 & 10.050 & 9.550 & 9.690 & 10.120 \\
SMMJ030236.15+000817.1 & 2.435 & 11.285 & 10.880 & 10.689 & 11.180 & 11.170 & 11.310 & 11.250 & 11.200 & 11.570 \\
SMMJ030238.62+001106.3 & 0.276 & 8.525 & 10.790 & 10.639 & 8.220 & 8.520 & 9.020 & 8.520 & 7.200 & 9.480 \\
SMMJ030244.82+000632.3 & 0.176 & 10.205 & 9.830 & 9.559 & 10.100 & 10.090 & 10.390 & 9.950 & 9.950 & 10.410 \\
SMMJ105151.69+572636.0 & 1.147 & 11.175 & 11.450 & 11.109 & 11.370 & 11.370 & 11.370 & 11.520 & 11.520 & 12.140 \\
SMMJ105155.47+572312.7 & 2.686 & 11.235 & 10.580 & 10.419 & 10.880 & 10.460 & 11.530 & 11.180 & 8.650 & 11.570 \\
SMMJ105158.02+571800.2 & 2.239 & 11.675 & 11.480 & 11.259 & 11.320 & 11.340 & 11.320 & 11.600 & 11.260 & 12.020 \\
SMMJ105200.22+572420.2 & 0.689 & 10.945 & 9.870 & 9.759 & 9.870 & 9.440 & 9.870 & 10.170 & 7.630 & 11.140 \\
SMMJ105201.25+572445.7 & 2.148 & 11.115 & 10.470 & 10.349 & 11.020 & 11.040 & 11.020 & 11.020 & 10.220 & 11.540 \\
SMMJ105207.49+571904.0 & 2.689 & 11.655 & 11.010 & 10.879 & 11.190 & 11.230 & 11.460 & 11.050 & 11.050 & 11.510 \\
SMMJ105225.79+571906.4 & 2.372 & 11.385 & 11.070 & 10.879 & 11.660 & 11.650 & 11.660 & 11.760 & 11.760 & 12.010 \\
SMMJ105227.58+572512.4 & 2.142 & 11.565 & 11.370 & 11.189 & 11.190 & 11.420 & 11.190 & 11.310 & 11.190 & 11.750 \\
SMMJ105227.77+572218.2 & 1.956 & 11.445 & 9.820 & 9.609 & -99.000 & -99.000 & 11.190 & -99.000 & -99.000 & -99.000 \\
SMMJ105230.73+572209.5 & 2.611 & 11.875 & 11.470 & 11.289 & 11.320 & 10.890 & 12.130 & 11.690 & 9.040 & 11.970 \\
SMMJ105238.19+571651.1 & 1.852 & 10.905 & 10.700 & 10.519 & 11.160 & 11.460 & 11.680 & 11.410 & 11.210 & 11.480 \\
SMMJ105238.30+572435.8 & 3.036 & 11.975 & 11.450 & 11.289 & 11.180 & 10.820 & 11.180 & 11.500 & 9.000 & 11.820 \\
SMMJ123549.44+621536.8 & 2.203 & 11.425 & 11.320 & 11.149 & 10.820 & 10.830 & 10.820 & 10.880 & 10.880 & 11.520 \\
SMMJ123553.26+621337.7 & 2.098 & 11.075 & 10.650 & 10.489 & 10.640 & 11.290 & 11.500 & 10.950 & 10.770 & 11.310 \\
SMMJ123555.14+620901.7 & 1.875 & 11.505 & 11.230 & 11.069 & 11.260 & 11.740 & 12.090 & 11.550 & 11.150 & 11.900 \\
SMMJ123600.10+620253.5 & 2.710 & 11.285 & -99.000 & -99.000 & 9.440 & 9.300 & 9.440 & 9.720 & 9.000 & 12.150 \\
SMMJ123600.15+621047.2 & 1.994 & 11.385 & 10.470 & 10.249 & 11.140 & 11.520 & 11.650 & 11.410 & 11.630 & 11.840 \\
SMMJ123606.72+621550.7 & 2.416 & 11.195 & 10.450 & 10.369 & 11.010 & 11.500 & 11.650 & 11.330 & 10.810 & 11.720 \\
SMMJ123606.85+621021.4 & 2.509 & 11.735 & 11.270 & 11.149 & 11.370 & 11.500 & 11.720 & 11.340 & 11.340 & 11.750 \\
SMMJ123616.15+621513.7 & 2.578 & 11.715 & 10.760 & 10.799 & 11.280 & 10.860 & 11.280 & 11.580 & 9.040 & 12.050 \\
SMMJ123618.33+621550.5 & 1.865 & 11.445 & 10.880 & 10.659 & 10.770 & 11.340 & 11.630 & 11.090 & 10.900 & 11.430 \\
SMMJ123621.27+621708.4 & 1.988 & 11.655 & 10.780 & 10.619 & 10.890 & 11.460 & 11.760 & 11.210 & 10.710 & 11.570 \\
SMMJ123622.65+621629.7 & 2.466 & 11.665 & 11.250 & 11.059 & 11.020 & 11.600 & 11.890 & 11.340 & 11.210 & 11.720 \\
SMMJ123629.13+621045.8 & 1.013 & 11.445 & 11.030 & 10.909 & 10.870 & 11.150 & 11.500 & 11.140 & 10.810 & 11.530 \\
SMMJ123632.61+620800.1 & 1.993 & 11.075 & 10.520 & 10.359 & 11.060 & 10.670 & 11.060 & 11.370 & 8.830 & 11.740 \\
SMMJ123634.51+621241.0 & 1.219 & 11.405 & 10.940 & 10.769 & 10.850 & 10.500 & 10.850 & 11.180 & 8.610 & 11.520 \\
SMMJ123635.59+621424.1 & 2.005 & 11.125 & 11.420 & 11.239 & 11.610 & 12.270 & 12.470 & 11.920 & 11.600 & 12.270 \\
SMMJ123636.75+621156.1 & 0.557 & 10.215 & 9.930 & 9.779 & 9.720 & 10.030 & 10.360 & 10.040 & 8.480 & 11.040 \\
SMMJ123651.76+621221.3 & 0.298 & 9.875 & -99.000 & -99.000 & 9.430 & 9.590 & 9.940 & 9.510 & 8.650 & 10.230 \\
SMMJ123707.21+621408.1 & 2.484 & 11.695 & 11.260 & 11.139 & 11.050 & 11.090 & 11.050 & 11.370 & 10.570 & 11.750 \\
SMMJ123711.98+621325.7 & 1.992 & 11.035 & 10.610 & 10.419 & 10.560 & 11.150 & 11.420 & 10.880 & 10.560 & 11.230 \\
SMMJ123712.05+621212.3 & 2.914 & 11.845 & 11.340 & 11.139 & 11.030 & 11.660 & 11.800 & 11.350 & 11.480 & 11.680 \\
SMMJ123716.01+620323.3 & 2.037 & 11.675 & -99.000 & -99.000 & 12.010 & 11.850 & 13.150 & 12.100 & 10.000 & 13.100 \\
SMMJ123721.87+621035.3 & 0.979 & 11.185 & 10.800 & 10.689 & 10.720 & 11.130 & 11.530 & 10.910 & 10.780 & 11.780 \\
SMMJ131201.17+424208.1 & 3.405 & 11.105 & 11.010 & 10.909 & 11.190 & 10.860 & 11.830 & 11.460 & 10.310 & 11.870 \\
SMMJ131208.82+424129.1 & 1.544 & 11.245 & 10.980 & 10.809 & 10.770 & 10.470 & 10.770 & 11.060 & 9.800 & 11.440 \\
SMMJ131212.69+424422.5 & 2.805 & 11.565 & 10.710 & 10.649 & 10.970 & 10.960 & 10.970 & 10.830 & 10.830 & 11.420 \\
SMMJ131215.27+423900.9 & 2.565 & 12.065 & 11.980 & 11.869 & 12.530 & 12.100 & 12.530 & 12.820 & 10.290 & 13.220 \\
SMMJ131222.35+423814.1 & 2.565 & 11.505 & 11.490 & 11.309 & 12.120 & 11.620 & 12.120 & 12.470 & 9.790 & 12.760 \\
SMMJ131225.20+424344.5 & 1.038 & 10.795 & 10.810 & 10.609 & 10.590 & 10.170 & 10.590 & 10.910 & 8.340 & 11.770 \\
SMMJ131225.73+423941.4 & 1.554 & 10.965 & 10.580 & 10.469 & 10.410 & 11.180 & 11.340 & 10.690 & 11.000 & 11.550 \\
SMMJ131228.30+424454.8 & 2.931 & 11.415 & 11.040 & 10.819 & 11.080 & 11.230 & 11.730 & 11.370 & 10.610 & 11.770 \\
SMMJ131231.07+424609.0 & 2.713 & 10.995 & -99.000 & -99.000 & 15.890 & 15.330 & 13.360 & 12.320 & 12.320 & 12.450 \\
SMMJ131232.31+423949.5 & 2.320 & 11.765 & 10.180 & 10.089 & 12.180 & 12.450 & 12.800 & 12.460 & 12.200 & 12.720 \\
SMMJ131239.14+424155.7 & 2.242 & 11.255 & 10.700 & 10.529 & 10.530 & 11.020 & 11.140 & 10.780 & 10.380 & 11.340 \\
SMMJ141741.81+522823.0 & 1.150 & 11.755 & 11.770 & 11.619 & 11.890 & 11.460 & 11.890 & 12.200 & 10.990 & 12.560 \\
SMMJ141742.04+523025.7 & 0.661 & 10.865 & 10.750 & 10.569 & 10.660 & 10.230 & 10.660 & 10.960 & 8.410 & 11.420 \\
SMMJ141750.50+523101.0 & 2.128 & 10.845 & 10.010 & 9.829 & 9.900 & 10.360 & 10.940 & 10.010 & 9.610 & 11.040 \\
SMMJ141800.40+512820.3 & 1.913 & 12.185 & 11.160 & 10.999 & 12.450 & 13.620 & 12.450 & 12.740 & 11.790 & 12.740 \\
SMMJ141802.87+523011.1 & 2.127 & 10.725 & 10.260 & 10.069 & 10.260 & 10.460 & 10.810 & 10.440 & 10.440 & 10.960 \\
SMMJ141809.00+522803.8 & 2.712 & 11.975 & 11.770 & 11.479 & 11.370 & 11.930 & 12.050 & 11.540 & 12.090 & 12.140 \\
SMMJ141813.54+522923.4 & 3.484 & 11.055 & 10.360 & 10.499 & 10.440 & 11.100 & 11.110 & 10.460 & 10.850 & 11.160 \\
SMMJ163627.94+405811.2 & 3.180 & 10.985 & 10.580 & 10.249 & 10.420 & 10.760 & 11.090 & 10.590 & 10.730 & 11.120 \\
SMMJ163631.47+405546.9 & 2.283 & 11.435 & 10.830 & 10.699 & 10.860 & 10.420 & 10.860 & 11.180 & 8.610 & 11.660 \\
SMMJ163639.01+405635.9 & 1.495 & 10.975 & 10.700 & 10.539 & 11.210 & 10.330 & 11.210 & 11.220 & 8.510 & 11.520 \\
SMMJ163650.43+405734.5 & 2.378 & 11.315 & 11.160 & 10.939 & 11.580 & 11.590 & 11.580 & 11.720 & 11.720 & 12.180 \\
SMMJ163658.19+410523.8 & 2.454 & 11.515 & 11.110 & 10.959 & 11.050 & 11.750 & 11.910 & 11.350 & 11.700 & 11.760 \\
SMMJ163658.78+405728.1 & 1.190 & 11.235 & 10.990 & 10.839 & 10.800 & 10.380 & 10.800 & 11.110 & 8.960 & 11.480 \\
SMMJ163704.34+410530.3 & 0.840 & 10.595 & 9.940 & 9.759 & 9.980 & 9.950 & 10.540 & 10.210 & 9.410 & 10.620 \\
SMMJ163706.51+405313.8 & 2.374 & 11.655 & 11.210 & 11.079 & 11.460 & 11.230 & 12.010 & 11.670 & 10.530 & 12.090 \\
SMMJ221724.69+001242.1 & 0.510 & 11.055 & -99.000 & -99.000 & 10.580 & 10.150 & 11.680 & 10.890 & 8.340 & 11.900 \\
SMMJ221725.97+001238.9 & 3.094 & 11.705 & -99.000 & -99.000 & 10.860 & 11.480 & 11.470 & 11.170 & 11.380 & 11.510 \\
SMMJ221733.02+000906.0 & 0.926 & 11.385 & -99.000 & -99.000 & 12.000 & 12.000 & 12.000 & 12.100 & 12.090 & 11.760 \\
SMMJ221733.12+001120.2 & 0.652 & 11.075 & 10.700 & 10.579 & 10.420 & 10.550 & 10.420 & 10.740 & 10.140 & 11.680 \\
SMMJ221733.91+001352.1 & 2.555 & 11.715 & 11.160 & 10.959 & 11.200 & 11.520 & 11.860 & 11.490 & 11.090 & 11.890 \\
SMMJ221735.15+001537.2 & 3.098 & 11.115 & 10.710 & 10.459 & 10.620 & 11.050 & 11.280 & 10.900 & 10.970 & 11.280 \\
SMMJ221735.84+001558.9 & 3.089 & 11.155 & 10.860 & 10.689 & 10.930 & 11.350 & 11.720 & 11.220 & 11.070 & 11.600 \\
SMMJ221737.39+001025.1 & 2.614 & 11.315 & -99.000 & -99.000 & 10.960 & 11.470 & 11.370 & 11.010 & 11.410 & 11.490 \\
SMMJ221804.42+002154.4 & 2.517 & 11.035 & 10.920 & 10.739 & 10.800 & 10.450 & 11.470 & 11.080 & 10.230 & 11.500 \\
SMMJ221806.77+001245.7 & 3.623 & 11.315 & 11.010 & 10.839 & 11.330 & 11.730 & 11.990 & 11.600 & 11.570 & 12.140 \\
\hline
\end{tabular}
\tablefoot{ 
The \citet{chabrier03} IMF is assumed. -99 indicates the lack of an estimate for a given SMG. 
\tablefoottext{a}{Stellar mass derived by \citetalias{michalowski10smg}.}
\tablefoottext{b}{Stellar mass derived by \citetalias{hainline11} with BC03 models.}
\tablefoottext{c}{Stellar mass derived by \citetalias{hainline11} with M05 models.}
\tablefoottext{d}{Stellar mass derived as in C10 with BC03 models and single-burst SFH.}
\tablefoottext{e}{Stellar mass derived as in C10 with BC03 models and tau SFH.}
\tablefoottext{f}{Stellar mass derived as in C10 with BC03 models and double-burst SFH.}
\tablefoottext{g}{Stellar mass derived as in C10 with M05 models and single-burst SFH.}
\tablefoottext{h}{Stellar mass derived as in C10 with M05 models and tau SFH.}
\tablefoottext{i}{Stellar mass derived as in C10 with M05 models and double-burst SFH.}
References: \citetalias{michalowski10smg} \citep{michalowski10smg}; \citetalias{hainline11} \citep{hainline11}; \citetalias{cirasuolo10} \citep{cirasuolo10}; BC03 \citep{bruzualcharlot03}; M05 \citep{maraston05}.
}
\end{table*}

%----------------------------------REFERENCES----------------------------------

%\input{/Users/michal/bibtex/getbibaa.tex}

\end{document}